\documentclass[12pt,letterpaper]{article}
\usepackage{amssymb}
\usepackage{amsmath}
\usepackage{amscd}
\usepackage{graphicx}
\usepackage{tikz}

\def\vCent#1{\vcenter{\hbox{\hss#1\hss}}}

\def\be{\begin{equation}}
\def\ee{\end{equation}}
\def\bea{\begin{eqnarray}}
\def\eea{\end{eqnarray}}
\def\bse{\begin{subequations}}
\def\ese{\end{subequations}}

\def\fracm#1#2{\hbox{\large{${\frac{{#1}}{{#2}}}$}}}

\def\pa{\partial}      
\newcommand{\bm}[1]{\mbox{\boldmath$#1$}}

\def\a{{\alpha}}
\def\b{{\beta}}
\def\g{{\gamma}}

\def\ad{{\dot{\alpha}}}
\def\bd{{\dot{\beta}}}
\def\gd{{\dot{\gamma}}}
\def\D{{\rm D}}
\def\Dd{{\bar{\rm D}}}

\def\[{\left[}
\def\]{\right]}

\font\ro=cmsy10                          
\def\kcr{{\hbox{\ro \char'170}}}                
\def\ktl{{\hbox{\ro \char'170}}}        
\def\ktr{{\hbox{\ro \char'170}}}        
\def\kbl{{\hbox{\ro \char'170}}}        
\def\kbr{{\hbox{\ro \char'170}}}        

\parskip=\medskipamount                 
\thicklines                         

\newskip\humongous \humongous=0pt plus 1000pt minus 1000pt
\def\caja{\mathsurround=0pt}
\def\eqalign#1{\,\vcenter{\openup2\jot \caja
        \ialign{\strut \hfil$\displaystyle{##}$&$
        \displaystyle{{}##}$\hfil\crcr#1\crcr}}\,}
\newif\ifdtup

\def\border{                                            
        \setlength{\unitlength}{1mm}
        \newcount\xco
        \newcount\yco
        \xco=-21
        \yco=12
        \begin{picture}(140,0)
        \put(\xco,\yco){$\ktl$}
        \advance\yco by-1
        {\loop
        \put(\xco,\yco){$\kcr$}
        \advance\yco by-2
        \ifnum\yco>-240
        \repeat
        \put(\xco,\yco){$\kbl$}}
        \xco=158
        \yco=12
        \put(\xco,\yco){$\ktr$}
        \advance\yco by-1
        {\loop
        \put(\xco,\yco){$\kcr$}
        \advance\yco by-2
        \ifnum\yco>-240
        \repeat
        \put(\xco,\yco){$\kbr$}}
        \put(-20,13){\tiny **University of Maryland * Center for String and
         Particle  Theory* Physics Department***University of Maryland *Center
        for String and Particle  Theory** }
        \put(-20,-241.5){\tiny **University of Maryland * Center for String and
         Particle  Theory* Physics Department***University of Maryland *Center
        for String and Particle  Theory** }
        \end{picture}
        \par\vskip-8mm}

\def\headpic{                                           
        \indent
        \setlength{\unitlength}{.4mm}
        \thinlines
        \par
        \begin{picture}(29,16)
        \put(165,16){\line(1,0){4}}
        \put(170,16){\line(1,0){4}}
        \put(180,16){\line(1,0){4}}
        \put(175,0){\line(1,0){4}}
        \put(180,0){\line(1,0){4}}
        \put(185,0){\line(1,0){4}}
        \put(169,0){\line(0,1){16}}
        \put(170,0){\line(0,1){16}}
        \put(179,0){\line(0,1){16}}
        \put(180,0){\line(0,1){16}}
        \put(184,0){\line(0,1){16}}
        \put(185,0){\line(0,1){16}}
        \put(169,16){\oval(8,32)[bl]}
        \put(170,16){\oval(8,32)[br]}
        \put(179,0){\oval(8,32)[tl]}
        \put(185,0){\oval(8,32)[tr]}
        \end{picture}
        \par\vskip-6.5mm
        \thicklines}
\def\endtitle{\end{quotation}\newpage}                  

\begin{document}

\border\headpic {\hbox to\hsize{March 2011 \hfill
{UMDEPP 11-003}}}
\par
{$~$ \hfill
{MIT-CTP-4220}}
\par
{$~$ \hfill
{hep-th/1103.3565}
}
\par

\setlength{\oddsidemargin}{0.3in}
\setlength{\evensidemargin}{-0.3in}
\begin{center}
\vglue .10in
{\large\bf A Codicil To Massless Gauge Superfields of\\
Higher Integer Superspins
 \footnote
{Supported in part  by National Science Foundation Grant
PHY-.}\  }
\\[.5in]

S.\, James Gates, Jr.\footnote{gatess@wam.umd.edu}
and K. Koutrolikos\footnote{koutrol@umd.edu}
\\[0.2in]

{\it Center for String and Particle Theory\\
Department of Physics, University of Maryland\\
College Park, MD 20742-4111 USA}\\[1.8in]

{\bf ABSTRACT}\\[.01in]
\end{center}
\begin{quotation}
{We study theories of 4D, ${\cal N}=1$ supersymmetric massless, arbitrary
integer superspins. A new state-of-the art is being established by the discovery of
a new series of such theories for arbitrary superspin $Y$ ($Y=s$ for any integer $s$)
The lowest member of the series is surprisingly found to be a previously established
formulation of the spin (3/2, 1) supermultiplet.}

${~~~}$ \newline
\endtitle


\section{Introduction}

~~~The current state-of-the-art understanding on the subject of 4D, $\cal N $ $=$ 1 integer higher 
spin supersymmetric multiplets was established in a work by Kuzenko and Sibiryakov \cite{Off2} 
(KS) wherein they gave two such formulations for each and every possible value of the integer 
superspin $Y$. These formulations are based on the introduction of constrained compensating 
superfields.   This seminal work laid a foundation for a number of latter studies \cite{Related}.
The goal of this work is to re-examine these schemes in order to be able to reproduce 
their results and, if possible, to discover new formulations in the case of integer superspins. This is 
exactly what will happen in the following. Their results will emerge naturally from our algorithm as 
a possible way a theory of higher, integer massless superspins can be formulated. Also we will 
discover the KS description is not the only consistent formulation and at least one alternative exists.

We approach the problem from a different angle, by studying actions for spinorial superfields.
This is supported from the naive observation that for a massless, integer higher superspin theory the
higher spin projection operator acting on the main superfield of the theory must give rise to a chiral
object with an even number of indices.
\be
\left(\Pi\Psi\right)_{\a(2s-1)}\propto \D^{\a_{2s}}\ \Dd^2\D_{(\a_{2s}}\pa^{\ad_1}{}_{\a_{2s-1}}
\dots\pa^{\ad_{s-1}}{}_{\a_{s+1}}\Psi_{\a(s))\ad(s-1)}  ~~. \label{eq01}
\ee
Therefore the main superfield of the theory must have an odd number of indices.   Hence 
the construction of a higher integer superspin theory must be developed around a spinorial 
superfield $\Psi_{\a(s)\ad(s-1)}$.



\section{The Setup}
~~As our starting point, we assume the main object of the theory to be a spinorial superfield 
$\Psi_{\a(s)\ad(s-1)}$, this means it's highest spin component (the $\theta\bar{\theta}$ term) 
must be a propagating fermion.   Therefore the mass dimensions of $\Psi$ must be $1/2$.  
The most general action which is quadratic for a spinorial superfield $\Psi_{\a(s)\ad(s-1)}$ with 
mass dimensions $\[\Psi\]=1/2$ has to include exactly 2 $\D$'s ($\Dd$'s) and thus takes the 
form:
\be
\eqalign{
S=\int d^8z \Big\{& \, 
c_1\Psi^{\a(s)\ad(s-1)}\D^2\Psi_{\a(s)\ad(s-1)}+c.c.\cr
&+c_2\Psi^{\a(s)\ad(s-1)}\Dd^2\Psi_{\a(s)\ad(s-1)} +c.c. \cr
&+\,a_1\Psi^{\a(s)\ad(s-1)}\Dd^{\ad_s}\D_{\a_s}{\Bar{\Psi}}_{\a(s-1)\ad(s)}\cr
&+a_2\Psi^{\a(s)\ad(s-1)}\D_{\a_s}\Dd^{\ad_s}{\Bar {\Psi}}_{\a(s-1)\ad(s)}\Big\}  ~~.
} \label{eq02} 
\ee

Since this action is going to describe a massless supermultiplet, it should be invariant under 
a gauge symmetry.  The most general gauge symmetry allowed by the invariance of the higher 
spin projection operator \eqref{eq01} has the following structure:
\be
\delta\Psi_{\a(s)\ad(s-1)}=\[\frac{1}{s!}\]\D_{(\a_s}K_{\a(s-1))\ad(s-1)}+\[\frac{1}{(s-1)!}\]\D_{
(\ad_{s-1}}\Lambda_{\a(s)\ad(s-2))}  ~~.  \label{eq03}
\ee
The change of  the above action with respect to the gauge transformation is:
\be
\eqalign{
\delta S=\int d^8z \Bigg\{&\left(-2c_1\D_{\a_s}\Psi^{\a(s)\ad(s-1)}+a_2\Dd_{\ad_s}\bar{\Psi
}^{\a(s-1)\ad(s)}\right)\D^{\b}\Dd_{\ad_{s-1}}\Lambda_{\b\a(s-1)\ad(s-2)}\cr
&+2c_2\Psi^{\a(s)\ad(s-1)}\Dd^2\D_{\a_s}K_{\a(s-1)\ad(s-1)} 
-a_1\bar{\Psi}^{\a(s-1)\ad(s)}D^2\Dd_{\ad_s}K_{\a(s-1)\ad(s-1)}\cr
&+\left(a_2\[\frac{s+1}{s}\]-a_1\right)\bar{\Psi}^{\a(s-1)\ad(s)}\Dd_{\ad_s}\D^2K_{\a(s-1)\ad(s-1)}\cr
&+a_1\[\frac{s-1}{s}\]\bar{\Psi}^{\a(s-1)\ad(s)}\D_{\a_{s-1}}\Dd_{\ad_s}\D^{\b}K_{\b\a(s-2)\ad(s-1)}
+c.c.\Bigg\}  ~~.
}
\label{eq04}
\ee

This expression will play a key role in discovering the different ways an integer higher spin 
superspin theory can be realized. Our ultimate goal is, based on the above equations 
\eqref{eq03} and \eqref{eq04}, to find all possible ways to build a gauge invariant action which 
on-shell has exactly the degrees of freedom to form a massless irreducible representation 
of the super Poincare group.
 
Following this path suggests the special case of $s=1$ has to be treated 
separately since the index structure drastically changes\footnote{The structure of the $\Lambda$ 
term in the gauge transformation \eqref{eq03} has to change and equation \eqref{eq04}
\\ $~~~~~~$ is 
simplified considerably.}. At this point we focus on the $s>1$ case


\section{Consideration for ${\bm s>1}$ Case}
~~~Now all we have to do is to find and introduce a set of appropiate and unconstraint 
compensators that will serve a double purpose. First of all they must give rise to a gauge 
invariant action and secondly this action on-shell must generate an irreducible representation 
of the super Poincare group. The second requirement, which is very restrictive, can be phrased 
in a different way. We can interpret it to say the invariant action constructed out of the superfield $\Psi$ 
and a set of compensators when expanded in component fields must give the massless
integer spin Fronsdal action for its bosonic piece and the massless half integer Fronsdal 
action for its fermionic piece.  Since $\Psi_{\a(s)\ad(s-1)}$ doesn't include all the fields 
needed for the Fronsdal actions, we need at least one propagating compensator.  In 
principle it can be either a bosonic superfield\footnote{This must possess a mass dimensions 
0.} or a fermionic superfield\footnote{This possesses mass dimensions 1/2.}.

Any attempt to introduce a fermionic compensator which has less indices than the main 
superfield appears precluded by the gauge invariance of the action. In order to succeed 
with the fermionic compensator it must be constrained\footnote{The constraint can be 
solved in terms of an unconstrained bosonic superfield.}. Therefore the propagating 
compensator needed has to be a bosonic one $V$ with zero mass dimensions. This 
follows the pattern of the higher spin theories developed so far in that the statistics flip
between the main superfield and the propagating prepotential

One more very important observation is  that $V$, must have a gauge transformation 
that involves either $K$ or $\Lambda$ parameters, since these are the only two available. 
But these parameters also have mass dimensions 0. That means that the gauge 
transformation will be algebraic (no derivatives are present). Ignoring the index structure 
and coefficients this gauge variation must look as
\be
\delta V \sim K + \Lambda  ~~~. \nonumber
\ee
~~~This is not acceptable because it means that $V$ can be {\em {completely}} gauged away 
and therefore there is nothing around to provide the extra degrees of freedom needed in order 
to form on-shell an irreducible multiplet. The only way out is if we allow the gauge parameters 
$K$ or $\Lambda$ to have some more $\D$-structure within them. In this way we could 
introduce a zero mass dimensions bosonic compensator with a gauge transformation 
which is not algebraic and therefore could be used to gauge away all the unwanted 
degrees of freedom. 

The last piece of information that we need in order to construct the full theory, is what type 
of gauge transformations for $V$, should we introduce? The structure of the Fronsdal action 
comes to the rescue. In the massless integer spin Fronsdal action, there are two real 
bosonic component fields:
\be
\eqalign{
&\text{the main field}~~ h_{\a(s)\ad(s)},~~~~\[h_{\a(s)\ad(s)}\]=1,~~~\delta h_{\a(s)\ad(s)}=
\[\frac{1}{s!^2}\]\pa_{(\a_s(\ad_s}\zeta_{\a(s-1))\ad(s-1))}~~, \cr  \nonumber
&\text{a compensator}~ h_{\a(s-2)\ad(s-2)},~\[h_{\a(s-2)\ad(s-2)}\]=1,~~\delta 
h_{\a(s-2)\ad(s-2)}=\pa^{\ad_s\a_s}\zeta_{\a(s-1)\ad(s-1)} ~.
}
\ee
The main superfield $\Psi$ can provide a component field with the index structure of $h_{\a(s)
\ad(s)}$, but not one for the role of $h_{\a(s-2)\ad(s-2)}$. This field has to come from the 
compensator $V$. So on-shell $V$ must provide only one real bosonic component with 
the proper index structure, mass dimensions and gauge transformation in order to play 
the role of $h_{\a(s-2)\ad(s-2)}$. This suggest that:
\begin{itemize}
\item $V$ should be real and therefore it's index structure must be $V_{\a(s-1)\ad(s-1)}$,

\item The $V^{(0,0)}_{\a(s-1)\ad(s-1)}$ component must be able to be gauged away, it 
has wrong mass dimensions and index structure. This can be achieved if:
\be 
\delta V_{\a(s-1)\ad(s-1)}|  \sim \text{~some component of the gauge parameter 
(algebraically)},  
  \nonumber
 \ee

\item The $V^{(2,0)}_{\a(s-1)\ad(s-1)}$ must be able to be gauged away (wrong index 
structure)
\be
\D^2 \delta V_{\a(s-1)\ad(s-1)}|\sim \text{~some component of the gauge parameter 
(algebraically),}\nonumber
\ee

\item The $V^{(1,1)(S,S)}_{\a(s)\ad(s)}$ component must be able to be gauged away\\
(wrong index structure)
\be \eqalign{
\[\D_{(\a_s},\Dd_{(\ad_s}\]\delta V_{\a(s-1))\ad(s-1))}| &\sim \text{~some component of 
the gauge parameter }  \cr
& \text{~~~~ (algebraically), ~and} 
 \nonumber
} \ee
 
 \item The $V^{(1,1)(A,A)}_{\a(s-2)\ad(s-2)}$ component must survive on-shell and 
 transform like
\be \eqalign{
\[\D^{\a_{s-1}},\Dd^{(\ad_{s-1}}\]\delta V_{\a(s-1)\ad(s-1)}|&\sim \pa^{\ad_{s-1}\a_{s-1}}\text{~
some component of the gauge}   \cr  
& \text{~~~~\,parameter.}   \nonumber
} \ee
 
\end{itemize}
 
These requirements fix the desired gauge transformation for the bosonic compensator 
to the following form:
\be
\delta V_{\a(s-1)\ad(s-1)}=\D^{\a_s}U_{\a(s)\ad(s-1)}+\Dd^{\ad_s}\bar{U}_{\a(s-1)\ad(s)}
~~.
\label{eq05}
\ee
 
Now it is very clear for what we are searching. By studying equation \eqref{eq04}, we 
will explore all possible ways that we can introduce a real bosonic compensator with
the above gauge transformation \eqref{eq05}. This can happen only by a couple of ways. 
The first thing that is clear is that the parameter $\Lambda$, because of it's index structure 
and the way it appears in \eqref{eq04}, can not have an internal structure such it that will 
lead to the introduction of the desired compensator. So our efforts must focus on the 
parameter $K$. That also means that if we insist on having a $\Lambda$-term in \eqref{eq03} 
we must introduce another compensator which must be auxiliary\footnote{in order not to 
introduce new degrees of freedom} or the coefficients related to the $\Lambda$ terms must 
vanish. 

\subsection{The KS-series}

~~~~By observing \eqref{eq04} we see that if $K_{\a(s-1)\ad{s-1}}=\D^{\a_s}U_{\a(s)\ad
(s-1)}$ then the last two terms vanish and the change of the action becomes
\be
\eqalign{
\delta S=\int d^8z &\left(-2c_1\D_{\a_s}\Psi^{\a(s)\ad(s-1)}+a_2\Dd_{\ad_s}\bar{\Psi}^{\a
(s-1)\ad(s)}\right)\D^{\b}\Dd_{\ad_{s-1}}\Lambda_{\b\a(s-1)\ad(s-2)}\cr
&+\left(2c_2\D_{\a_s}\Dd^2\Psi^{\a(s)\ad(s-1)}-a_1\Dd_{\ad_s}\D^2\bar{\Psi}^{\a(s-1)
\ad(s)}\right)\D^{\b}U_{\b\a(s-1)\ad(s-1)}\cr
&+c.c.
} \label{equ06}
\ee
Now it is obvious that if
\be
\eqalign{
-2c_1&=a_2   ~~,~~
2c_2=-a_1  ~~~,
}     \label{equ07}
\ee
the variation in (\ref{equ06}) becomes
\be
\eqalign{
\delta S=\int d^8z\ &a_2\left(\D_{\a_s}\Psi^{\a(s)\ad(s-1)}+\Dd_{\ad_s}\bar{\Psi}^{\a(s-1
)\ad(s)}\right)\left[\D^{\b}\Dd_{\ad_{s-1}}\Lambda_{\b\a(s-1)\ad(s-2)}+c.c.\right]\cr
-&a_1\left(\D_{\a_s}\Dd^2\Psi^{\a(s)\ad(s-1)}+\Dd_{\ad_s}\D^2\bar{\Psi}^{\a(s-1)
\ad(s)}\right)\left[\D^{\b}U_{\b\a(s-1)\ad(s-1)}+c.c.\right]
~~. }   \label{equ08}
\ee

The observations above suggest the introduction of two real compensators\\
$B_{\a(s-1)\ad(s-1)},~V_{\a(s-1)\ad(s-1)}$ with mass dimensions $[B]=1,~[V]=0$ and 
gauge transformations
\be
\eqalign{
&\delta B_{\a(s-1)\ad(s-1)}=\[\frac{1}{(s-1)!}\]\Big(\D^{\a_s}\Dd_{(\ad_{s-1}}\Lambda_{
\a(s)\ad(s-2))}+\Dd^{\ad_s}\D_{(\a_{s-1}}\bar{\Lambda}_{\a(s-2))\ad(s)}\Big) ~~, \cr
&\delta V_{\a(s-1)\ad(s-1)}=\D^{\a_s}U_{\a(s)\ad(s-1)}+\Dd^{\ad_s}\bar{U}_{\a(s-1)\ad(s)}
~~. }    \label{equ09}
\ee
The gauge transformation of $\Psi$ becomes
\be
\delta\Psi_{\a(s)\ad(s-1)}=-\D^2U_{\a(s)\ad(s-1)}+\[\frac{1}{(s-1)!}\]\Dd_{(\ad_{s-1}}
\Lambda_{\a(s)\ad(s-2))}  ~~.       \label{equ10}
\ee
In order to construct a gauge invariant action and the compensators to have dynamics 
we add the following terms in the action:

\begin{itemize}

\item Counter terms (they cancel the change of the initial action)
\be
\eqalign{
S_{c}=\int d^8z \Bigg\{-&a_2\left(\D_{\a_s}\Psi^{\a(s)\ad(s-1)}+\Dd_{\ad_s}\bar{\Psi}^{
\a(s-1)\ad(s)}\right)B_{\a(s-1)\ad(s-1)}\cr
+&a_1\left(\D_{\a_s}\Dd^2\Psi^{\a(s)\ad(s-1)}+\Dd_{\ad_s}\D^2\bar{\Psi}^{\a(s-1)\ad(s)}\right)
V_{\a(s-1)\ad(s-1)}\Bigg\}  ~~,
}   \label{equ11}
\ee

\item Kinetic energy terms (the most general free action for each of the compensators)
\be
\eqalign{
S_{k.e}=\int d^8z~&eB^{\a(s-1)\ad(s-1)}B_{\a(s-1)\ad(s-1)}\cr
+&h_1 V^{\a(s-1)\ad(s-1)}\D^{\g}\Dd^2\D_{\g} V_{\a(s-1)\ad(s-1)}\cr
+&h_2 V^{\a(s-1)\ad(s-1)}\Box V_{\a(s-1)\ad(s-1)}\cr
+&h_3V^{\a(s-1)\ad(s-1)}\pa_{\a_{s-1}\ad_{s-1}}\pa^{\gd\g} V_{\g\a(s-2)\gd\ad(
s-2)}\cr
+&h_4 V^{\a(s-1)\ad(s-1)}\left[\D_{\a_{s-1}},\Dd_{\ad_{s-1}}\right]\left[\D^{\g},
\Dd^{\gd}\right] V_{\g\a(s-2)\gd\ad(s-2)}  ~~,
}   \label{equ12}
\ee

\item Interaction terms (in principle there might be interactions among compensators)
\be
\eqalign{
S_{int.}=\int d^8z bB^{\a(s-1)\ad(s-1)}\left(\D^2V_{\a(s-1)\ad(s-1)}+\Dd^2V_{\a(s-1)\ad(
s-1)}\right) ~~. }     \label{equ13}
\ee

\end{itemize}

Thus the full action look as below and contains only a series of constants to be determined.
\be
\eqalign{  {~~~~~}
S=\int d^8z\Bigg\{
-&\frac{1}{2}a_2\Psi^{\a(s)\ad(s-1)}\D^2\Psi_{\a(s)\ad(s-1)}+c.c.\cr
-&\frac{1}{2}a_1\Psi^{\a(s)\ad(s-1)}\Dd^2\Psi_{\a(s)\ad(s-1)} +c.c. \cr
+&\,a_1\Psi^{\a(s)\ad(s-1)}\Dd^{\ad_s}\D_{\a_s}{\Bar {\Psi}}_{\a(s-1)\ad(s)}\cr
+&a_2\Psi^{\a(s)\ad(s-1)}\D_{\a_s}\Dd^{\ad_s}{\Bar {\Psi}}_{\a(s-1)\ad(s)}\cr
-&a_2\left(\D_{\a_s}\Psi^{\a(s)\ad(s-1)}+\Dd_{\ad_s}\bar{\Psi}^{\a(s-1)\ad(s)}\right)
B_{\a(s-1)\ad(s-1)}\cr
+&a_1\left(\D_{\a_s}\Dd^2\Psi^{\a(s)\ad(s-1)}+\Dd_{\ad_s}\D^2\bar{\Psi}^{\a(s-1)
\ad(s)}\right)V_{\a(s-1)\ad(s-1)}\cr
+&eB^{\a(s-1)\ad(s-1)}B_{\a(s-1)\ad(s-1)}\cr
+&h_1V^{\a(s-1)\ad(s-1)}\D^{\g}\Dd^2\D_{\g}V_{\a(s-1)\ad(s-1)}\cr
+&h_2V^{\a(s-1)\ad(s-1)}\Box V_{\a(s-1)\ad(s-1)}\cr
+&h_3V^{\a(s-1)\ad(s-1)}\pa_{\a_{s-1}\ad_{s-1}}\pa^{\gd\g}V_{\g\a(s-2)\gd\ad(s-2)}\cr
+&h_4V^{\a(s-1)\ad(s-1)}\left[\D_{\a_{s-1}},\Dd_{\ad_{s-1}}\right]\left[\D^{\g},\Dd^{\gd}
\right]V_{\g\a(s-2)\gd\ad(s-2)}\cr
+&bB^{\a(s-1)\ad(s-1)}\left(\D^2V_{\a(s-1)\ad(s-1)}+\Dd^2V_{\a(s-1)\ad(s-1)}\right)\Bigg\}  ~~.
}  \label{equ14}
\ee
The requirement that this action is invariant under the above transformations, will give 
rise to two Bianchi identities required for the gauge invariance of the action.
\be
\eqalign{
&\D^2{\bm{\cal T}}_{\a(s)\ad(s-1)}+\[\frac{1}{s!}\]\D_{(\a_s}{\bm{\cal P}}_{\a(s-1))
\ad(s-1)}=0 ~~, \cr
&\Dd^{\ad_{s-1}}{\bm{\cal T}}_{\a(s)\ad(s-1)}-\[\frac{1}{s!}\]\Dd^{\ad_{s-1}}D_{
(\a_s}{\bm{\cal G}}_{\a(s-1))\ad(s-1)}=0 ~~,
}   \label{equ15}
\ee
where ${\bm{\cal T}}_{\a(s)\ad(s-1)},~{\bm{\cal P}}_{\a(s-1)\ad(s-1)},~{\bm{\cal G}}_{\a(s-1)
\ad(s-1)}$ are the variations of the action with respect to the superfields $\Psi_{\a(s)\ad(s-1)
},~V_{\a(s-1)\ad(s-1)},~B_{\a(s-1)\ad(s-1)}$.  The solution of the first one gives:
\be
\eqalign{
&h_1=\frac{1}{2}a_1 ~~~,~~~ h_2=0~~~,~~
h_3=0 ~~~,~~~h_4=0  ~~~,~~~
b=0  ~~~,
}     \label{equ16}
\ee
and the second one gives:
\be
\eqalign{
&e=-\frac{1}{2}a_2 ~~~,~~~ b=0  ~~~. \cr
}     \label{equ17}
\ee

So the gauge invariant action is:
\be
\eqalign{
S=\int d^8z\Bigg\{
-&\frac{1}{2}a_2\Psi^{\a(s)\ad(s-1)}\D^2\Psi_{\a(s)\ad(s-1)}+c.c.\cr
-&\frac{1}{2}a_1\Psi^{\a(s)\ad(s-1)}\Dd^2\Psi_{\a(s)\ad(s-1)} +c.c. \cr
+&\,a_1\Psi^{\a(s)\ad(s-1)}\Dd^{\ad_s}\D_{\a_s}{\Bar {\Psi}}_{\a(s-1)\ad(s)}\cr
+&a_2\Psi^{\a(s)\ad(s-1)}\D_{\a_s}\Dd^{\ad_s}{\Bar {\Psi}}_{\a(s-1)\ad(s)}\cr
-&a_2\left(\D_{\a_s}\Psi^{\a(s)\ad(s-1)}+\Dd_{\ad_s}\bar{\Psi}^{\a(s-1)\ad(s)}
\right)B_{\a(s-1)\ad(s-1)}\cr
+&a_1\left(\D_{\a_s}\Dd^2\Psi^{\a(s)\ad(s-1)}+\Dd_{\ad_s}\D^2\bar{\Psi}^{
\a(s-1)\ad(s)}\right)V_{\a(s-1)\ad(s-1)}\cr
-&\frac{1}{2}a_2B^{\a(s-1)\ad(s-1)}B_{\a(s-1)\ad(s-1)}\cr
+&\frac{1}{2}a_1V^{\a(s-1)\ad(s-1)}\D^{\g}\Dd^2\D_{\g}V_{\a(s-1)\ad(s-1)}
\Bigg\}  ~~~.
}    \label{equ18}
\ee

Now we can integrate out the auxiliary superfield $B$.  Using the on-shell equation of 
motion of $B_{\a(s-1)\ad(s-1)}$ and substitute it back in to the action we get:
\be
\eqalign{
S=\int d^8z\Bigg\{
-&\frac{1}{2}a_1\Psi^{\a(s)\ad(s-1)}\Dd^2\Psi_{\a(s)\ad(s-1)} +c.c. \cr
+&\,a_1\Psi^{\a(s)\ad(s-1)}\Dd^{\ad_s}\D_{\a_s}{\Bar {\Psi}}_{\a(s-1)\ad(s)}\cr
+&a_1\left(\D_{\a_s}\Dd^2\Psi^{\a(s)\ad(s-1)}+\Dd_{\ad_s}\D^2\bar{\Psi}^{\a(s-1)\ad(s)}\right)
V_{\a(s-1)\ad(s-1)}\cr
+&\frac{1}{2}a_1V^{\a(s-1)\ad(s-1)}\D^{\g}\Dd^2\D_{\g}V_{\a(s-1)\ad(s-1)}\Bigg\}
~~, }     \label{equ19}
\ee
and this action is invariant under the transformations
\be
\eqalign{
&\delta\Psi_{\a(s)\ad(s-1)}=-\D^2U_{\a(s)\ad(s-1)}+\[\frac{1}{(s-1)!}\]\Dd_{(\ad_{s-1
}}\Lambda_{\a(s)\ad(s-2))}\cr
&\delta V_{\a(s-1)\ad(s-1)}=\D^{\a_s}U_{\a(s)\ad(s-1)}+\Dd^{\ad_s}\bar{U}_{\a(s-1)\ad(s)}
~~. }     \label{equ20}
\ee
This theory\footnote{We could have reached the same result if from the very begging we 
have choosen \\ $~~~~~~$ $c_1=a_2=0$ instead of introducing the auxiliary compensator 
$B$.} is equivalent to the theory of  S. Kuzenko and A. Sibiryakov \cite{Off2}, once one solves 
the constraints that appear in their description (as done in \cite{Gates:2010td}).
This theory is well studied and it is known to describe on-shell  a massless supermultiplet of superspin $Y$=$s$.

\subsection{The FVdWH-series}

~~~~Again by observing equation \eqref{eq04} we find that there is another way to arrange 
things. By setting $K_{\a(s-1)\ad(s-1)}=\Dd^{\ad_s}U_{\a(s-1)\ad(s)}$ then we find:
\be
\eqalign{
\delta S=\int d^8z &\left(-2c_1\D_{\a_s}\Psi^{\a(s)\ad(s-1)}+a_2\Dd_{\ad_s}\bar{\Psi}^{
\a(s-1)\ad(s)}\right)\D^{\b}\Dd_{\ad_{s-1}}\Lambda_{\b\a(s-1)\ad(s-2)}\cr
&+2c_2\Psi^{\a(s)\ad(s-1)}\Dd^2\D_{\a_s}\Dd^{\ad_s}U_{\a(s-1)\ad(s)} 
-a_1\bar{\Psi}^{\a(s-1)\ad(s)}D^2\Dd_{\ad_s}\Dd^{\ad_s}U_{\a(s-1)\ad(s)}\cr
&+\left(a_2\[\frac{s+1}{s}\]-a_1\right)\bar{\Psi}^{\a(s-1)\ad(s)}\Dd_{\ad_s}\D^2
\Dd^{\ad_s}U_{\a(s-1)\ad(s)}\cr
&+a_1\[\frac{s-1}{s}\]\bar{\Psi}^{\a(s-1)\ad(s)}\D_{\a_{s-1}}\Dd_{\ad_s}\D^{\b}
\Dd^{\ad_s}U_{\b\a(s-2)\ad(s)}\cr
&+c.c.
}      \label{equ21}
\ee
and this suggest setting 
\be
a_2\[\frac{s+1}{s}\]=a_1  ~~,    \label{equ22}
\ee
so that we find
\be
\eqalign{
\delta S=\int d^8z &\left(-2c_1\D_{\a_s}\Psi^{\a(s)\ad(s-1)}+a_2\Dd_{\ad_s}\bar{
\Psi}^{\a(s-1)\ad(s)}\right)\D^{\b}\Dd_{\ad_{s-1}}\Lambda_{\b\a(s-1)\ad(s-2)}\cr
&+\left(2c_2\D_{\a_s}\Dd^2\Psi^{\a(s)\ad(s-1)}-a_1\Dd_{\ad_s}\D^2\bar{\Psi}^{
\a(s-1)\ad(s)}\right)\Dd^{\bd}U_{\a(s-1)\bd\ad(s-1)}\cr
&-a_1\[\frac{s-1}{s}\]\D^{\a_{s-1}}\Dd_{\bd}\D_{\b}\bar{\Psi}^{\b\a(s-2)\bd\ad(s-1)}\left(
\Dd^{\ad_s}U_{\a(s-1)\ad(s)}+c.c.\right)\cr
&+c.c.
}      \label{equ23}
\ee
In order to minimise the degrees of freedom that we have to introduce and construct 
a minimal theory we set:
\be
\eqalign{
-2c_1&=a_2  ~~,~~
2c_2=-a_1  ~~,
}     \label{equ24}
\ee
so the change of the action takes the form
\be
\eqalign{
\delta S=\int d^8z ~&a_2\D_{\a_s}\Psi^{\a(s)\ad(s-1)}\[\frac{1}{(s-1)!}\]\[\D^{\b}
\Dd_{(\ad_{s-1}}\Lambda_{\b\a(s-1)\ad(s-2))}+c.c.\]\cr
+&a_2\Dd_{\ad_s}\bar{\Psi}^{\a(s-1)\ad(s)}\[\frac{1}{(s-1)!}\]\left[\D^{\b}\Dd_{
(\ad_{s-1}}\Lambda_{\b\a(s-1)\ad(s-2))}+c.c.\right]\cr
-&a_1\D_{\a_s}\Dd^2\Psi^{\a(s)\ad(s-1)}\left[\Dd^{\bd}U_{\a(s-1)\bd\ad(s-1)}
+c.c.\right]\cr
-&a_1\Dd_{\ad_s}\D^2\bar{\Psi}^{\a(s-1)\ad(s)}\left[\Dd^{\bd}U_{\a(s-1)\bd\ad
(s-1)}+c.c.\right]\cr
-&a_1\[\frac{s-1}{s}\]\left(\D^{\a_{s-1}}\Dd_{\bd}\D_{\b}\bar{\Psi}^{\b\a(s-2
)\bd\ad(s-1)}+c.c.\right)\left[\Dd^{\ad_s}U_{\a(s-1)\ad(s)}+c.c.\right]   ~~~.
}     \label{equ25}
\ee

We introduce two real compensators, $B_{\a(s-1)\ad(s-1)},~V_{\a(s-1)\ad(s-1)}$
with $[B]=1,~[V]=0$ and gauge transformations
\be
\eqalign{
&\delta B_{\a(s-1)\ad(s-1)}=\[\frac{1}{(s-1)!}\]\left[\D^{\a_s}\Dd_{(\ad_{s-1}}
\Lambda_{\a(s)\ad(s-2))}+\Dd^{\ad_s}\D_{(\a_{s-1}}\bar{\Lambda}_{\a(s-2
))\ad(s)}\right]  ~~, \cr
&\delta V_{\a(s-1)\ad(s-1)}=\Dd^{\ad_s}U_{\a(s-1)\ad(s)}+\D^{\a_s}\bar{
U}_{\a(s)\ad(s-1)}   ~~,
}      \label{equ26}
\ee
and the gauge transformation of the $\Psi$ superfield is
\be
\delta\Psi_{\a(s)\ad(s-1)}=\[\frac{1}{s!}\]\D_{(\a_s}\Dd^{\ad_s}U_{\a(s-1))
\ad(s)}+\[\frac{1}{(s-1)!}\]\Dd_{(\ad_{s-1}}\Lambda_{\a(s)\ad(s-2))}     ~~~. 
\label{equ27}
\ee 
Hence we have to add a few terms to the action

\begin{itemize}

\item Counter terms (they cancel the change of the initial action)
\be
\eqalign{
S_c=\int d^8z\Big\{-&\[\frac{s}{s+1}\]a_1\left(\D_{\a_s}\Psi^{\a(s)\ad(s-1)}+\Dd_{\ad_s}
\bar{\Psi}^{\a(s-1)\ad(s)}\right)B_{\a(s-1)\ad(s-1)}\cr
+&a_1\left(\D_{\a_s}\Dd^2\Psi^{\a(s)\ad(s-1)}+\Dd_{\ad_s}\D^2\bar{\Psi}^{\a(s-1)\ad(s)
}\right)V_{\a(s-1)\ad(s-1)}\cr
+&a_1\[\frac{s-1}{s}\]\left(\Dd^{\ad_{s-1}}\D_{\b}\Dd_{\bd}\Psi^{\b\a(s-1)\bd\ad(s-2)}+
c.c.\right)V_{\a(s-1)\ad(s-1)} ~~,
}     \label{equ28}
\ee

\item Kinetic energy (the most general action for each of the compensators)
\be
\eqalign{
S_{k.e}=\int d^8z
~&eB^{\a(s-1)\ad(s-1)}B_{\a(s-1)\ad(s-1)}\cr
+&h_1V^{\a(s-1)\ad(s-1)}\D^{\g}\Dd^2\D_{\g}V_{\a(s-1)\ad(s-1)}\cr
+&h_2V^{\a(s-1)\ad(s-1)}\Box V_{\a(s-1)\ad(s-1)}\cr
+&h_3V^{\a(s-1)\ad(s-1)}\pa_{\a_{s-1}\ad_{s-1}}\pa^{\g\gd}V_{\g\a(s-2)\gd\ad(s-2)}\cr
+&h_4V^{\a(s-1)\ad(s-1)}\left[\D_{\a_{s-1}},\Dd_{\ad_{s-1}}\right]\left[\D^{\g},\Dd^{\gd}
\right]V_{\g\a(s-2)\gd\ad(s-2)}  ~~,
}       \label{equ29}
\ee

\item Interaction terms (in principle there might be interactions among compensators)
\be
S_{int}=\int d^8z bB^{\a(s-1)\ad(s-1)}\left(\D^2V_{\a(s-1)\ad(s-1)}+\Dd^2V_{\a(s-1)
\ad(s-1)}\right)  ~~.       \label{equ30}
\ee
\end{itemize}

Therefore the full action is
\be
\eqalign{  {~~~~~}
S=\int d^8z \Big\{
-&\frac{1}{2}\[\frac{s}{s+1}\]a_1\Psi^{\a(s)\ad(s-1)}\D^2\Psi_{\a(s)\ad(s-1)}+c.c.\cr
-&\frac{1}{2}a_1\Psi^{\a(s)\ad(s-1)}\Dd^2\Psi_{\a(s)\ad(s-1)} +c.c. \cr
+&a_1\Psi^{\a(s)\ad(s-1)}\Dd^{\ad_s}\D_{\a_s}{\Bar{\Psi}}_{\a(s-1)\ad(s)}\cr
+&\[\frac{s}{s+1}\]a_1\Psi^{\a(s)\ad(s-1)}\D_{\a_s}\Dd^{\ad_s}{\Bar {\Psi}}_{\a(s-1
)\ad(s)}\cr
-&\[\frac{s}{s+1}\]a_1\left(\D_{\a_s}\Psi^{\a(s)\ad(s-1)}+\Dd_{\ad_s}\bar{\Psi}^{\a(s-1
)\ad(s)}\right)B_{\a(s-1)\ad(s-1)}\cr
+&a_1\left(\D_{\a_s}\Dd^2\Psi^{\a(s)\ad(s-1)}+\Dd_{\ad_s}\D^2\bar{\Psi}^{\a(s-1)
\ad(s)}\right)V_{\a(s-1)\ad(s-1)}\cr
+&\[\frac{s-1}{s}\]a_1\left(\Dd^{\ad_{s-1}}\D_{\b}\Dd_{\bd}\Psi^{\b\a(s-1)\bd\ad(s-2
)}+c.c.\right)V_{\a(s-1)\ad(s-1)}\cr
+&eB^{\a(s-1)\ad(s-1)}B_{\a(s-1)\ad(s-1)}\cr
+&h_1V^{\a(s-1)\ad(s-1)}\D^{\g}\Dd^2\D_{\g}V_{\a(s-1)\ad(s-1)}\cr
+&h_2V^{\a(s-1)\ad(s-1)}\Box V_{\a(s-1)\ad(s-1)}\cr
+&h_3V^{\a(s-1)\ad(s-1)}\pa_{\a_{s-1}\ad_{s-1}}\pa^{\g\gd}V_{\g\a(s-2)\gd\ad(s-2)}\cr
+&h_4V^{\a(s-1)\ad(s-1)}\left[\D_{\a_{s-1}},\Dd_{\ad_{s-1}}\right]\left[\D^{\g},\Dd^{\gd}
\right]V_{\g\a(s-2)\gd\ad(s-2)}\cr
+&bB^{\a(s-1)\ad(s-1)}\left(\D^2V_{\a(s-1)\ad(s-1)}+\Dd^2V_{\a(s-1)\ad(s-1)}\right)\Big\}
~~~. }      \label{equ31}
\ee

The invariance of this action under the gauge transformations is guaranteed by the 
following two Bianchi Identities (derived as before):
\be
\eqalign{
\[\frac{1}{s!}\]\Dd_{(\ad_s}\D^{\a_s}{\bm{\cal T}}_{\a(s)\ad(s-1))}-\[\frac{1}{s!}\]\Dd_{(\ad_s}
{\bm{\cal P}}_{\a(s-1)\ad(s-1))}=0   ~~, \cr
\Dd^{\ad_{s-1}}{\bm{\cal T}}_{\a(s)\ad(s-1)}-\[\frac{1}{s!}\]\Dd^{\ad_{s-1}}\D_{(\a_s}{\bm{\cal 
G}}_{\a(s-1))\ad(s-1)}=0  ~~,
}     \label{equ32}
\ee
where ${\bm{\cal T}}_{\a(s)\ad(s-1)},~{\bm{\cal P}}_{\a(s-1)\ad(s-1)},~{\bm{\cal G}}_{\a
(s-1)\ad(s-1)}$ are the variations of the action with respect to the superfields $\Psi_{
\a(s)\ad(s-1)},~V_{\a(s-1)\ad(s-1)},~B_{\a(s-1)\ad(s-1)}$.

The solution of the first one gives:
\be
\eqalign{
&h_1=\[\frac{1}{2s}\]a_1 ~~~,~~~~~~~~~~~~\,~~~~~~~ h_2=-\[\frac{s-1}{2s}\]a_1 ~~,\cr
&h_3=\[\frac{(2s-1)(s-1)}{(2s)^2}\]a_1 ~~~,~~~ h_4=\[\frac{s-1}{(2s)^2}\]a_1 ~,\cr
&b=-a_1 ~~~,   
}   \label{equ33}
\ee
and the second one has as a solution:
\be
\eqalign{
&e=-\frac{1}{2}\[\frac{s}{s+1}\]a_1 ~~~,~~~ b=-a_1   ~~.
}    \label{equ34}
\ee

Thus the action takes the form
\be
\eqalign{ {~~~~~}
S=\int d^8z \Big\{
-&\frac{1}{2}\[\frac{s}{s+1}\]a_1\Psi^{\a(s)\ad(s-1)}\D^2\Psi_{\a(s)\ad(s-1)}+c.c.\cr
-&\frac{1}{2}a_1\Psi^{\a(s)\ad(s-1)}\Dd^2\Psi_{\a(s)\ad(s-1)} +c.c. \cr
+&a_1\Psi^{\a(s)\ad(s-1)}\Dd^{\ad_s}\D_{\a_s}{\Bar{\Psi}}_{\a(s-1)\ad(s)}\cr
+&\[\frac{s}{s+1}\]a_1\Psi^{\a(s)\ad(s-1)}\D_{\a_s}\Dd^{\ad_s}{\Bar {\Psi}}_{\a(s-1)
\ad(s)}\cr
-&\[\frac{s}{s+1}\]a_1\left(\D_{\a_s}\Psi^{\a(s)\ad(s-1)}+\Dd_{\ad_s}\bar{\Psi}^{\a(s-1
)\ad(s)}\right)B_{\a(s-1)\ad(s-1)}\cr
+&a_1\left(\D_{\a_s}\Dd^2\Psi^{\a(s)\ad(s-1)}+\Dd_{\ad_s}\D^2\bar{\Psi}^{\a(s-1)
\ad(s)}\right)V_{\a(s-1)\ad(s-1)}\cr
+&\[\frac{s-1}{s}\]a_1\left(\Dd^{\ad_{s-1}}\D_{\b}\Dd_{\bd}\Psi^{\b\a(s-1)\bd\ad(s-2)}
+c.c.\right)V_{\a(s-1)\ad(s-1)}\cr
-&\frac{1}{2}\[\frac{s}{s+1}\]a_1B^{\a(s-1)\ad(s-1)}B_{\a(s-1)\ad(s-1)}\cr
+&\frac{1}{2s}a_1V^{\a(s-1)\ad(s-1)}\D^{\g}\Dd^2\D_{\g}V_{\a(s-1)\ad(s-1)}\cr
-&\[\frac{s-1}{2s}\]a_1V^{\a(s-1)\ad(s-1)}\Box V_{\a(s-1)\ad(s-1)}\cr
+&\[\frac{(2s-1)(s-1)}{(2s)^2}\]a_1V^{\a(s-1)\ad(s-1)}\pa_{\a_{s-1}\ad_{s-1}}
\pa^{\g\gd}V_{\g\a(s-2)\gd\ad(s-2)}\cr
+&\[\frac{s-1}{(2s)^2}\]a_1V^{\a(s-1)\ad(s-1)}\left[\D_{\a_{s-1}},\Dd_{\ad_{s-1
}}\right]\left[\D^{\g},\Dd^{\gd}\right]V_{\g\a(s-2)\gd\ad(s-2)}\cr
-&a_1B^{\a(s-1)\ad(s-1)}\left(\D^2V_{\a(s-1)\ad(s-1)}+\Dd^2V_{\a(s-1)\ad(s-1
)}\right)\Big\}   ~~~.
}      \label{equ35}
\ee

At this point we can integrate out the auxiliary superfield $B$ by using it's equation 
of motion
\be
\eqalign{
B_{\a(s-1)\ad(s-1)}=~&\left[\D^{\a_s}\Psi_{\a(s)\ad(s-1)}+\Dd^{\ad_s}\bar{\Psi}_{\a(s-1
)\ad(s)}\right]\cr
-&\[\frac{s+1}{s}\]\left(\D^2V_{\a(s-1)\ad(s-1)}+\Dd^2V_{\a(s-1)\ad(s-1)}\right)  ~~,
}     \label{equ36}
\ee
and substituting this result back in (\ref{equ35}) yields
\be
\eqalign{
S=\int d^8z \Big\{
-&\frac{1}{2}a_1\Psi^{\a(s)\ad(s-1)}\Dd^2\Psi_{\a(s)\ad(s-1)} +c.c. \cr
+&a_1\Psi^{\a(s)\ad(s-1)}\Dd^{\ad_s}\D_{\a_s}{\Bar{\Psi}}_{\a(s-1)\ad(s)}\cr
+&a_1\left\{\Dd^2,\D_{\a_s}\right\}\Psi^{\a(s)\ad(s-1)}V_{\a(s-1)\ad(s-1)}+c.c.\cr
+&\[\frac{s-1}{s}\]a_1\Dd^{\ad_{s-1}}\D_{\b}\Dd_{\bd}\Psi^{\b\a(s-1)\bd\ad(s-2)}
V_{\a(s-1)\ad(s-1)}+c.c.\cr
+&\[\frac{s+2}{2s}\]a_1V^{\a(s-1)\ad(s-1)}\D^{\g}\Dd^2\D_{\g}V_{\a(s-1)\ad(s-1)}\cr
+&\[\frac{1}{s}\]a_1V^{\a(s-1)\ad(s-1)}\Box V_{\a(s-1)\ad(s-1)}\cr
+&\[\frac{(2s-1)(s-1)}{(2s)^2}\]a_1V^{\a(s-1)\ad(s-1)}\pa_{\a_{s-1}\ad_{s-1}}
\pa^{\g\gd}V_{\g\a(s-2)\gd\ad(s-2)}\cr
+&\[\frac{s-1}{(2s)^2}\]a_1V^{\a(s-1)\ad(s-1)}\left[\D_{\a_{s-1}},\Dd_{\ad_{s-1
}}\right]\left[\D^{\g},\Dd^{\gd}\right]V_{\g\a(s-2)\gd\ad(s-2)}\Big\}  ~~. \label{action}
}      
\ee

Calculating variations with respect to $\Psi$ and $V$ in this action we can define the following superfields
\be
\eqalign{
{\bm{\cal T}}_{\a(s)\ad(s-1)}=-&a_1\Dd^2\Psi_{\a(s)\ad(s-1)}+\frac{a_1}{s!}
\Dd^{\ad_s}\D_{(\a_s}\bar{\Psi}_{\a(s-1))\ad(s)}  {~~~~~~~~~~~~~~}  \cr
+&\frac{a_1}{s!}\Dd^2\D_{(\a_s}V_{\a(s-1))\ad(s-1)}+\frac{a_1}{s!}\D_{(\a_s}
\Dd^2V_{\a(s-1))\ad(s-1)}\cr
-&\frac{s-1}{s!s!}a_1\Dd_{(\ad_{s-1}}\D_{(\a_s}\Dd^{\bd}V_{\a(s-1))\bd\ad(s-2))}
~~, }    \label{equ38}
\ee

$$
\eqalign{ {~~~~~}
{\bm{\cal P}}_{\a(s-1)\ad(s-1)}=-&a_1\D^{\a_s}\Dd^2\Psi_{\a(s)\ad(s-1)}-a_1\Dd^{\ad_s}
\D^2\bar{\Psi}_{\a(s-1)\ad(s)}   {~~~~~~~~~~~~~~~~~~~~}  \cr
-&a_1\Dd^2\D^{\a_s}\Psi_{\a(s)\ad(s-1)}-a_1\D^2\Dd^{\ad_s}\bar{\Psi}_{\a(s-1)\ad(s)}\cr
+&a_1\frac{s-1}{s!}\Dd_{(\ad_{s-1}}\D^{\b}\Dd^{\bd}\Psi_{\b\a(s-1)\bd\ad(s-2))}\cr
+&a_1\frac{s-1}{s!}\D_{(\a_{s-1}}\Dd^{\bd}\D^{\b}\bar{\Psi}_{\b\a(s-2))\bd\ad(s-1)}\cr
+&a_1\frac{s+2}{s}\D^{\g}\Dd^2\D_{\g}V_{\a(s-1)\ad(s-1)}\cr
+&a_1\frac{2}{s}\Box V_{\a(s-1)\ad(s-1)}\cr
}   
$$
\be
\eqalign{ {~~~~~}
~~~~~~~~~~~~~~+&a_1\frac{(2s-1)(s-1)}{2s!^2}\pa_{(\a_{s-1}(\ad_{s-1}}\pa^{\b\bd}
V_{\b\a(s-2)) \bd\ad(s-2))}\cr
+&a_1\frac{s-1}{2s!^2}\left[\D_{(\a_{s-1}},\Dd_{(\ad_{s-1}}\right]\left[\D^{\b},
\Dd^{\bd}\right]V_{\b\a(s-2))\bd\ad(s-2))}  ~~, 
}    \label{equ39}
\ee
and they satisfy the Bianchi Identities for this final action
\be
\eqalign{
&\Dd^{\ad_{s-1}}{\bm{\cal T}}_{\a(s)\ad(s-1)}=0  ~~, \cr
&\frac{1}{s!}\Dd_{(\ad_s}\D^{\a_s}{\bm{\cal T}}_{\a(s)\ad(s-1))}-\frac{1}{s!}\Dd_{(\ad_s}
{\bm{\cal P}}_{\a(s-1)\ad(s-1))}=0   ~~.
}    \label{equ40}
\ee
Furthermore we can prove that they also satisfy the following identity:
\be
\eqalign{
\Dd^{\ad_{2s}}\bar{\bm{\cal W}}_{\ad(2s)}=&-i\frac{2s}{a_1}\pa^{\a_{s}}{}_{(\ad_{2s-1
}}\dots\pa^{\a_{1}}{}_{\ad_{s}}{\bm{\cal T}}_{\a(s)\ad(s-1))}\cr
&-\frac{s}{a_1}\Dd^2\pa^{\a_{s-1}}{}_{(\ad_{2s-1}}\dots\pa^{\a_{1}}{}_{\ad_{s+1}}
\Bar{\bm{\cal T}}_{\a(s-1)\ad(s))}\cr
&+\frac{s}{a_1}\Dd_{(\ad_{2s-1}}\pa^{\a_{s-1}}{}_{\ad_{2s-2}}\dots\pa^{\a_{1}}
{}_{\ad_{s}}{\bm{\cal P}}_{\a(s-1)\ad(s-1))}  ~~, 
}     \label{equ42}
\ee
where the anti-chiral superfield (i.e. $\D_{\b} \bar{\bm{\cal W}}_{\ad(2s)}= 0$) is given by
\be
\bar{\bm{\cal W}}_{\ad(2s)}=\D^2\Dd_{(\ad_{2s}}\pa^{\a_{s-1}}{}_{\ad_{2s-1}}\dots
\pa^{\a_{1}}{}_{\ad_{s+1}}\bar{\Psi}_{\a(s-1)\ad(s))}   ~~.   \label{equ42}
\ee
So this theory has an irreducible multiplet propagating on-shell with superspin $Y=s$. 
Now we can check if these are the only degrees of freedom propagating.\\

Expanding the superfields $\Psi,~V$ into components and using their gauge transformations 
we find that some components have purely algebraic transformations and therefore can 
be gauged away. In detail\footnote{The definition of symmetric and antisymmetric
pieces of a field is the following $$\Phi_{\g\a(s-1)}=\Phi^{(S)}_{\g\a(s-1)}+\frac{s-1}{s!}
 C_{\g(\a_{s-1}}\Phi^{(A)}_{\a(s-2))}~~,~~\Phi^{(S)}_{\g\a(s-1)}=\frac{1}{s!}\Phi_{(\g\a(s-1))}
 ~~,~~\Phi^{(A)}_{\a(s-2)}=C^{\g\a_{s-1}}\Phi_{\g\a(s-1)}$$
 Furthermore the notation $\Phi^{(m,n)}$ represents the $\theta^{m}\bar{\theta}^n$ 
 component in the taylor series of the superfield $\Phi$}:\vskip.2in
\begin{tabular}{c c}
For Bosons & {~~} For Fermions
\\
\begin{tabular}{| l | l |}
\hline
Component & Gauged away by\\
\hline
$V^{(0,0)}_{\a(s-1)\ad(s-1)}$ & $Re\left[U^{(0,1)(A)}_{\a(s-1)\ad(s-1)}\right]$\\
\hline
$V^{(1,1)(S,S)}_{\a(s)\ad(s)}$ & $Re\left[U^{(1,2)(S)}_{\a(s)\ad(s)}\right]$\\
\hline
$V^{(1,1)(A,S)}_{\a(s-2)\ad(s)}$ & $U^{(1,2)(A)}_{\a(s-2)\ad(s)}$\\
\hline
$V^{(2,0)}_{\a(s-1)\ad(s-1)}$ & $U^{(2,1)(A)}_{\a(s-1)\ad(s-1)}$\\
\hline
$\Psi^{(1,0)(S)}_{\a(s+1)\ad(s-1)}$ & $\Lambda^{(1,1)(S,S)}_{\a(s+1)\ad(s-1)}$\\ 
\hline
$\Psi^{(1,0)(A)}_{\a(s-1)\ad(s-1)}$ & $\Lambda^{(1,1)(A,S)}_{\a(s-1)\ad(s-1)}$\\
\hline
$Im\left[\Psi^{(0,1)(S)}_{\a(s)\ad(s)}\right]$ & $Im\left[U^{(1,2)(S)}_{\a(s)\ad(s)}\right]$\\
\hline
$\Psi^{(2,1)(A)}_{\a(s)\ad(s-2)}$ & $\Lambda^{(2,2)}_{\a(s)\ad(s-2)}$\\
\hline
\end{tabular}
& {~~~~}
\begin{tabular}{| l | l |}
\hline
Component & Gauged away by\\
\hline
$V^{(0,1)(S)}_{\a(s-1)\ad(s)}$ & $U^{(0,2)}_{\a(s-1)\ad(s)}$\\
\hline
$V^{(1,0)(A)}_{\a(s-2)\ad(s-1)}$ & $U^{(1,1)(A,A)}_{\a(s-2)\ad(s-1)}$\\
\hline
$V^{(2,1)(S)}_{\a(s-1)\ad(s)}$ & $U^{(2,2)}_{\a(s-1)\ad(s)}$\\
\hline
$\Psi^{(0,0)}_{\a(s)\ad(s-1)}$ & $\Lambda^{(0,1)(S)}_{\a(s)\ad(s-1)}$\\
\hline
$\Psi^{(2,0)}_{\a(s)\ad(s-1)}$ & $\Lambda^{(2,1)(S)}_{\a(s)\ad(s-1)}$\\
\hline
$\Psi^{(1,1)(S,A)}_{\a(s+1)\ad(s-2)}$ & $\Lambda^{(1,2)(S)}_{\a(s+1)\ad(s-2)}$\\
\hline
$\Psi^{(1,1)(A,A)}_{\a(s-1)\ad(s-2)}$ & $\Lambda^{(1,2)(A)}_{\a(s-1)\ad(s-2)}$\\
\hline
\end{tabular}
\end{tabular}

So in the Wess-Zumino gauge the two superfields take the forms
\be
\eqalign{
V_{\a(s-1)\ad(s-1)}&=\[\frac{2(s-1)}{(s-1)!^2}\]\theta_{(\a_{s-1}}\bar{\theta
}_{(\ad_{s-1}}h_{\a(s-2))\ad(s-2))}\cr
&~~+\[\frac{\sqrt{2}}{(s-1)!}\]\theta^2\bar{\theta}_{(\ad_{s-1}}\psi_{\a(s-1)\ad(s-2))}\cr
&~~-\[\frac{\sqrt{2}}{(s-1)!}\]\theta_{(\a_{s-1}}\bar{\theta}^2\bar{\psi}_{\a(s-2))\ad(s-1)}
+\theta^2\bar{\theta}^2P_{\a(s-1)\ad(s-1)}  ~~,   \label{taylorA}
}  
\ee
and
\be
\eqalign{
\Psi_{\a(s)\ad(s-1)}&=\bar{\theta}^{\ad_s}h_{\a(s)\ad(s)}+\sqrt{2}\bar{\theta}^2\psi_{
\a(s)\ad(s-1)}+\sqrt{2}\theta^{\a_{s+1}}\bar{\theta}^{\ad_{s}}\psi_{\a(s+1)\ad(s)}\cr
&~~~+\[\frac{1}{s!}\]\theta_{(\a_s}\bar{\theta}^{\ad_{s}}\[\lambda_{\a(s-1))\ad(s)}-\frac{
\sqrt{2}s}{s+1}\bar{\psi}_{\a(s-1))\ad(s)}\]\cr
&~~~+\theta^2\bar{\theta}^{\ad_s}Y_{\a(s)\ad(s)}\cr
&~~~+\bar{\theta}^2\theta^{\a_{s+1}}\[t_{\a(s+1)\ad(s-1)}+i\frac{3}{2(s+1)!}\pa_{(
\a_{s+1}}{}^{\ad_s}h_{\a(s))\ad(s)}\]\cr
&~~~+\[\frac{s}{(s+1)!}\]\bar{\theta}^2\theta_{(\a_{s}}\[M_{\a(s-1))\ad(s-1)}+\frac{(s+1
)^2}{s(2s+1)}P_{\a(s-1))\ad{s-1}}\]\cr
&~~~+i\[\frac{s}{(s+1)!}\]\bar{\theta}^2\theta_{(\a_{s}}\[N_{\a(s-1))\ad(s-1)}+\frac{2s-1}
{2}\pa^{\g\gd}h_{\g\a(s-1))\gd\ad(s-1)}\right.\cr
&~~~~~~~~~~~~~~~~~~~~~~~~~~~~~~~-\left.\frac{(s+1)^2(s-1)}{s!}\pa_{\a_{s-1}
(\ad_{s-1}}h_{\a(s-2))\ad(s-2))}\]\cr
&~~~+\theta^2\bar{\theta}^2\[\chi_{\a(s)\ad(s-1)}+\frac{i}{\sqrt{2}}\frac{s-1}{s+1}\pa^{
\a_{s+1}\ad_s}\psi_{\a(s+1)\ad(s)}\right.\cr
&~~~~-\left.\frac{i}{\sqrt{2}}\frac{s(s-1)}{(s+1)^2s!}\pa_{(\a_s}{}^{\ad_s}\bar{\psi}_{\a(
s-1))\ad(s)}-\frac{i}{2(s+1)!}\pa_{(\a_s}{}^{\ad_s}\lambda_{\a(s-1))\ad(s)}\]
~~,  \label{taylorM}
}
\ee 
where the components $t,~M,~N,~P,~Y,~\lambda,~\chi$ will be shown to be auxiliary
fields.  All the others are symmetric in all undotted and dotted indices separately and the 
components $h_{\a(s)\ad(s)},~h_{\a(s-2)\ad(s-2)}$ are real. The component 
action for all the bosons is:  
$$
\eqalign{
S_{Bosons}=\int d^4x \Big\{&-2a_1h^{\a(s)\ad(s)}\Box h_{\a(s)\ad(s)}  
{~~~~~~~~~~~~~~~~~~~~~~~~~~~~~~~~~~~~~~~~~~~~~~~~~~~~~~~~~~} \cr
&+sa_1h^{\a(s)\ad(s)}\pa_{\a_{s}\ad_{s}}\pa^{\gd\g}h_{\g\a(s-1)\gd\ad(s-1)}\cr
&-\left[2s(s-1)\right]a_1h^{\a(s)\ad(s)}\pa_{\a_s\ad_s}\pa_{\a_{s-1}\ad_{s-1}}
h_{\a(s-2)\ad(s-2)}\cr
&+\left[2s(2s-1)\right]a_1h^{\a(s-2)\ad(s-2)}\Box h_{\a(s-2)\ad(s-2)}\cr
}
$$
\be
\eqalign{
{~~~~~~~} &+\left[s(s-2)^2\right]a_1h^{\a(s-2)\ad(s-2)}\pa_{\a_{s-2}\ad_{s-2}}
\pa^{\gd\g} h_{\g\a(s-3)\gd\ad(s-3)}\cr
&+  \frac{1}{2}a_1t^{\a(s+1)\ad(s-1)}t_{\a(s+1)\ad(s-1)}+c.c.\cr
&+ a_1 \[\frac{2s+1}{s+1}\]M^{\a(s-1)\ad(s-1)}M_{\a(s-1)\ad(s-1)}+c.c.\cr
&-  a_1 \[\frac{(s+1)^3}{s^2(2s+1)}\]P^{\a(s-1)\ad(s-1)}P_{\a(s-1)\ad(s-1)}+c.c.\cr
&+ a_1 \[\frac{1}{s+1}\]N^{\a(s-1)\ad(s-1)}N_{\a(s-1)\ad(s-1)}+c.c.\cr
&-a_1Y^{\a(s)\ad(s)}\bar{Y}_{\a(s)\ad(s)}  ~~,
}  \label{equ45}
\ee
and using the equations of motion for the auxiliary fields
\be
\eqalign{
&M_{\a(s-1)\ad(s-1)}=0 ~~,~~~ 
N_{\a(s-1)\ad(s-1)}=0  ~~,~~~ 
P_{\a(s-1)\ad(s-1)}=0 ~~~, \cr
&t_{\a(s+1)\ad(s-1)}=0 ~~,~~~ Y_{\a(s)\ad(s)}=0 ~~~, \cr
}  \label{equ46}
\ee
we obtain
\be
\eqalign{
S_{Bosons}=\int d^4x \Big\{&-2a_1h^{\a(s)\ad(s)}\Box h_{\a(s)\ad(s)} 
+sa_1h^{\a(s)\ad(s)}\pa_{\a_{s}\ad_{s}}\pa^{\gd\g}h_{\g\a(s-1)\gd\ad(s-1)}\cr
&-\left[2s(s-1)\right]a_1h^{\a(s)\ad(s)}\pa_{\a_s\ad_s}\pa_{\a_{s-1}\ad_{s-1}}
h_{\a(s-2)\ad(s-2)}\cr
&+\left[2s(2s-1)\right]a_1h^{\a(s-2)\ad(s-2)}\Box h_{\a(s-2)\ad(s-2)}\cr
&+\left[s(s-2)^2\right]a_1h^{\a(s-2)\ad(s-2)}\pa_{\a_{s-2}\ad_{s-2}}
\pa^{\gd\g}h_{\g\a(s-3)\gd\ad(s-3)} ~~, 
}  \label{equ47}
\ee
and upon by setting $a_1=-\frac{1}{2}$ we find
\be
\eqalign{
S_{Bosons}=\int d^4x \Big\{&h^{\a(s)\ad(s)}\Box h_{\a(s)\ad(s)}
-\frac{s}{2}h^{\a(s)\ad(s)}\pa_{\a_{s}\ad_{s}}\pa^{\gd\g}h_{\g\a(s-1)\gd\ad(s-1)}\cr
&+\left[s(s-1)\right]h^{\a(s)\ad(s)}\pa_{\a_s\ad_s}\pa_{\a_{s-1}\ad_{s-1}}h_{\a(
s-2)\ad(s-2)}\cr
&-\left[s(2s-1)\right]h^{\a(s-2)\ad(s-2)}\Box h_{\a(s-2)\ad(s-2)}\cr
&-\left[\frac{s(s-2)^2}{2}\right]h^{\a(s-2)\ad(s-2)}\pa_{\a_{s-2}\ad_{s-2}}
\pa^{\gd\g}h_{\g\a(s-3)\gd\ad(s-3)}   ~~~,
} \label{equ48}
\ee
which is the Fronsdal action for a propagating massless spin-$s$ bosonic field.\\

The fermionic piece of the action is:
$$
\eqalign{
S_{Fermions}=\int d^4x\Big\{
&-i2a_1\bar{\psi}^{\a(s)\ad(s+1)}\pa^{\a_{s+1}}{}_{\ad_{s+1}}\psi_{\a(s+1)\ad(s)}
{~~~~~~~~~~~~~~~~~~~}  {~~~~~~~~~~~~~~~~~~}  \cr
&+i2\[\frac{2s+1}{(s+1)^2}\]a_1\bar{\psi}^{\a(s-1)\ad(s)}\pa^{\a_s}{}_{\ad_s}\psi_{
\a(s)\ad(s-1)}\cr
&+i2a_1\bar{\psi}^{\a(s-2)\ad(s-1)}\pa^{\a_{s-1}}{}_{\ad_{s-1}}\psi_{\a(s-1)\ad(s-2)}\cr
} 
$$
\be
\eqalign{
{~~~\,~} &-i2\[\frac{s}{s+1}\]a_1\psi^{\a(s+1)\ad(s)}\pa_{\a_{s+1}\ad_s}\psi_{\a(s)
\ad(s-1)}+c.c.\cr
&+i2a_1\psi^{\a(s)\ad(s-1)}\pa_{\a_s\ad_{s-1}}\psi_{\a(s-1)\ad(s-2)}+c.c.\cr
&+\[\frac{s+1}{s}\]\bar{\lambda}^{\a(s)\ad(s-1)}\chi_{\a(s)\ad(s-1)}+c.c.
}  \label{equ49}
\ee
so that using the equations of motions for the auxiliary fields $\chi,~\lambda$
\be
\lambda_{\a(s-1)\ad(s)}=0 ~~, ~~~~ \chi_{\a(s)\ad(s-1)}=0 ~~,      \label{equ50}
\ee
and setting the value of $a_1=-\frac{1}{2}$,we get the final fermionic action
\be
\eqalign{
S_{Fermions}=\int d^4x\Big\{
&i\bar{\psi}^{\a(s)\ad(s+1)}\pa^{\a_{s+1}}{}_{\ad_{s+1}}\psi_{\a(s+1)\ad(s)}\cr
&-i\[\frac{2s+1}{(s+1)^2}\]\bar{\psi}^{\a(s-1)\ad(s)}\pa^{\a_s}{}_{\ad_s}\psi_{\a(s)
\ad(s-1)}\cr
&-i\bar{\psi}^{\a(s-2)\ad(s-1)}\pa^{\a_{s-1}}{}_{\ad_{s-1}}\psi_{\a(s-1)\ad(s-2)}\cr
&+i\[\frac{s}{s+1}\]\psi^{\a(s+1)\ad(s)}\pa_{\a_{s+1}\ad_s}\psi_{\a(s)\ad(s-1)}+c.c.\cr
&-i\psi^{\a(s)\ad(s-1)}\pa_{\a_s\ad_{s-1}}\psi_{\a(s-1)\ad(s-2)}+c.c.
}   \label{equ51}
\ee
This is the Frondsal action for a propagating massles spin-$(s+1/2)$ 
Therefore we conclude that only an irreducible supermultiplet propagates on-shell 
and therefore the action \eqref{action} describes a massless integer superspin
$Y$ $=$ $s$. 

The counting of the off-shell bosonic and fermionic degrees of freedom for the
action including all the auxiliary fields is:
\begin{center}
\begin{tabular} {| c | c | c |}
\hline
Component Field(s) & Bosonic  & Fermionic\\
\hline
$h_{\a(s)\ad(s)}$~/~$h_{\a(s-2)\ad(s-2)}$ & $s^2  + 2$ & {}\\
\hline
$\psi_{\a(s+1)\ad(s)}$~/~$\psi_{\a(s)\ad(s-1)}$~/~$\psi_{\a(s-1)\ad(s-2)}$  & {} & $4 ( 
s^2  + s + 1 ) $\\
\hline
$P_{\a(s-1)\ad(s-1)}$ & $s^2$ & {}\\
\hline
$\lambda_{\a(s-1)\ad(s)}$ & {} & $2 s (s+1) $\\
\hline
$Y_{\a(s)\ad(s)} $ & $ 2(s+1)^2$ & {}\\
\hline
$t_{\a(s+1)\ad(s-1)} $ & $2(s+2)s$ & {}\\
\hline
$
M_{\a(s-1)\ad(s-1)}$ & $s^2$ & {}\\
\hline
$N_{\a(s-1)\ad(s-1)}$ & $s^2$ & {}\\
\hline
$\chi_{\a(s)\ad(s-1)}$ & {} & $2 s (s+1) $\\
\hline\hline
$~$ & $8s^2+8s+4$ & $8s^2+8s+4$\\
\hline
\end{tabular}
\end{center}
For each case we have verified the existence 
of field strength superfields ${\bm{\cal W}}_{\a(2s)}$, $ {\bm{{\cal P}}}_{\a(s-1)\ad(
s-1)}$ and ${\bm{{\cal T}}}_{\a(s)\ad(s-1)}$ which occur for both the KS-series and the
FVdWH-series.

\section{Considering the $s = 1$ case}

~~~For the special case of $s=1$, the most general action takes the form:
\be
\eqalign{
S=\int d^8z \Big\{
&\, ~a_1\Psi^{\a}\Dd^{\ad}\D_{\a}{\Bar{\Psi}}_{\ad}
+a_2\Psi^{\a}\D_{\a}\Dd^{\ad}{\Bar {\Psi}}_{\ad} \cr
&~+ \big[ c_1\Psi^{\a}\D^2\Psi_{\a} 
~+c_2\Psi^{\a}\Dd^2\Psi_{\a} +c.c.  \big]  \Big\}   ~~.
} \label{equ52}
\ee

Also in this case, the gauge parameter $\Lambda_{\a(s)\ad(s-2)}$, in order to survive,
must be modified to $\Dd^{\ad_{s-1}}\Lambda_{\a(s)\ad(s-1)}$. So the most general 
gauge transformation allowed in the $s=1$ is:
\be
\delta\Psi_{\a}=\D_{\a}K+\Dd^2\Lambda_{\a}     ~~~.  \label{equ53}
\ee
The change of the general action (\ref{equ52}) under this transformation is:
\be
\eqalign{
\delta S=\int d^8z &\left(-2c_1\D_{\a}\Psi^{\a}+a_2\Dd_{\ad}\bar{\Psi}^{\ad}\right)\D^{\b}
\Dd^2\Lambda_{\b}+c.c.\cr
&+2c_2\Psi^{\a}\Dd^2\D_{\a}K+c.c.\cr
&-a_1\bar{\Psi}^{\ad}\D^2\Dd_{\ad}K+c.c.\cr
&+\left(2a_2-a_1\right)\bar{\Psi}^{\ad}\Dd_{\ad}\D^2K+c.c.\cr
} \label{equ54} 
\ee

At this point, we can do a very useful observation. All the propagating degrees of freedom 
required for the formulation of a massless $Y=1$ theory, can be included in the main 
superfield $\Psi_{\a}$.  To verify that, just look the Taylor expansion of the superfields 
\eqref{taylorM} and \eqref{taylorA}.  So as a consequence, either we have to make the
 change of the action to vanish or add purely auxiliary compensators.

\subsection{A) ~~$K=\D^{\a}U_{\a}$}
For this choice of $K$ we find the following action
\be
\eqalign{
S=\int d^8z\Bigg\{
-&\frac{1}{2}a_1\Psi^{\a}\Dd^2\Psi_{\a} +c.c. \cr
+&a_1\Psi^{\a}\Dd^{\ad}\D_{\a}{\Bar {\Psi}}_{\ad}\cr
+&a_1\left(\D_{\a}\Dd^2\Psi^{\a}+\Dd_{\ad}\D^2\bar{\Psi}^{\ad}\right)V\cr
+&\frac{1}{2}a_1V\D^{\g}\Dd^2\D_{\g}V\Bigg\}~~,
} \label{equ55}
\ee
and this action is invariant under the transformations
\be
\eqalign{
&\delta\Psi_{\a}=-\D^2U_{\a}+\Dd^2\Lambda_{\a}~~,\cr
&\delta V=\D^{\a}U_{\a}+\Dd^{\ad}\bar{U}_{\ad}~~,
}  \label{equ56}
\ee
and the Bianchi Identities are:
\be
\eqalign{
&\D^2{\bm{\cal T}}_{\a}+\D_{\a}{\bm{\cal P}}=0~~,\cr
&\Dd^2{\bm{\cal T}}_{\a}=0~~.
}  \label{equ57}
\ee
where
\be
\eqalign{
&{\bm{\cal T}}_{\a}=-a_1\Dd^2\Psi_{\a}+a_1\Dd^{\ad}\D_{\a}\bar{\Psi}_{\ad}+a_1
\Dd^2\D_{\a}V~~,\cr
&{\bm{\cal P}}=a_1\D^{\g}\Dd^2\D_{\g}V-a_1\left(\D^{\a}\Dd^2\Psi_{\a}+\Dd^{\ad}
\D^2\bar{\Psi}_{\ad}\right)~~.\cr
}   \label{equ58}
\ee
This is the $s=1$ limit of \eqref{equ19}.

On-shell the propagating degrees of freedom of superfield $V$ are gauged away completely 
and the only thing that survives is the $Y=1$ supermultiplet. This can be visualized by the 
following argument. There is a gauge where $V=0$. This happens for $U_{\a}=i\Dd^2\D_{\a}
L$. Working in this gauge the $V$ superfield vanishes from the action which becomes:
\be
\eqalign{
S=\int d^8z\Bigg\{
-&\frac{1}{2}a_1\Psi^{\a}\Dd^2\Psi_{\a} -\frac{1}{2}a_1\bar{\Psi}^{\ad}\D^2\bar{\Psi}_{\ad}  +a_1\Psi^{\a}\Dd^{\ad}\D_{\a}{\Bar {\Psi}}_{\ad}
\Bigg\}~~,
} \label{equ59}
\ee
and is invariant under the transformation
\be
\delta\Psi_{\a}=\D^2\Dd^{\ad}\pa_{\a\ad}L+\Dd^2\Lambda_{\a}~~.
\label{equ60}
\ee
This is the formulation suggested first by Fradkin and Vasiliev in \cite{Fradkin:1979as} and de 
Wit and van Holten in \cite{deWvanH} at the component level and  in \cite{Gates:1979gv} for a 
superfield description.  This last work also made the observation that this formulation was distinct
from an earlier off-shell description of the $Y$ $=$ 1 supermultiplet \cite{Ogievetsky:1975vk}.

\subsection{B) ~~$K=\Dd^{\ad}U_{\ad}$}
For this choice of $K$ we get the action:
\be
\eqalign{
S=\int d^8z \Big\{
-&\frac{1}{2}a_1\Psi^{\a}\Dd^2\Psi_{\a} +c.c. \cr
+&a_1\Psi^{\a}\Dd^{\ad}\D_{\a}{\Bar{\Psi}}_{\ad}\cr
+&a_1\left(\D_{\a}\Dd^2\Psi^{\a}+\Dd_{\ad}\D^2\bar{\Psi}^{\ad}\right)V\cr
+&a_1\left(\Dd^2\D_{\a}\Psi^{\a}+\D^2\Dd_{\ad}\bar{\Psi}^{\ad}\right)V\cr
+&\frac{3}{2}a_1V\D^{\g}\Dd^2\D_{\g}V\cr
+&a_1V\Box V\Big\}~~,
} \label{equ61}
\ee
and it is invariant under the gauge transformations
\be
\eqalign{
&\delta\Psi_{\a}=\D_{\a}\Dd^{\ad}U_{\ad}+\Dd^2\Lambda_{\a}~~,\cr
&\delta V=\Dd^{\ad}U_{\ad}+\D^{\a}\bar{U}_{\a} ~~,
}  \label{equ62}
\ee
with Bianchi Identities
\be
\eqalign{
&\Dd_{\ad}\D^{\a}{\bm{\cal T}}_{\a}-\Dd_{\ad}{\bm{\cal P}}=0\cr
&\Dd^2{\bm{\cal T}}_{\a}=0
}  \label{equ63}
\ee
with
\be
\eqalign{
&{\bm{\cal T}}_{\a}=-a_1\Dd^2\Psi_{\a}+a_1\Dd^{\ad}\D_{\a}\bar{\Psi}_{\ad}+a_1\left\{\Dd^2,\D_{\a}\right\}V~~,\cr
&{\bm{\cal P}}=3a_1\D^{\g}\Dd^2\D_{\g}V+2a_1\Box V-a_1\left[\left\{\Dd^2,\D^{\a}\right\}\Psi_{\a}+\left\{\D^2,\Dd^{\ad}\right\}\bar{\Psi}_{\ad}\right]~~.\cr
}  \label{equ64}
\ee
Like before there is no component of $V$ surviving on-shell. 
The only  propagating sub-multiplet is the $Y=1$ supermultiplet. This is the $s=1$
limit of  \eqref{action}.

\subsection{C) One more thing...\\
${~~~~}K=\bar{K},~\Lambda_{\a}=i\D_{\a}U,~U=\bar{U}$}
For the special case of $s=1$ there is one more possibility.\\
If $K=\bar{K}$,~$\Lambda_{\a}=i\D_{\a}U$,~$U=\bar{U}$ the change of the
action becomes
 \be
\eqalign{
\delta S=\int d^8z &~ i\left(-2c_1-a_2\right)\D_{\a}\Psi^{\a}\D^{\b}\Dd^2\D_{\b}U+c.c.\cr
&+\left(2c_2-a_1\right)\Psi^{\a}\Dd^2\D_{\a}K+c.c.\cr
&+\left(2a_2-a_1\right)\bar{\Psi}^{\ad}\Dd_{\ad}\D^2K+c.c.\cr
}    \label{equ65}
\ee
which suggests that by choosing:
\be
\eqalign{
-2c_1&=a_2  ~~,~~
2c_2 =a_1~~,~~
2a_2 =a_1  ~~.
}   \label{equ66}
\ee
we get:
\be
\eqalign{
S=\int d^8z \Big\{
&+a_1\Psi^{\a}\Dd^{\ad}\D_{\a}{\Bar{\Psi}}_{\ad}
+\fracm{1}{2}a_1\Psi^{\a}\D_{\a}\Dd^{\ad}{\Bar {\Psi}}_{\ad}\cr
&~+ a_1 \big[ ~ - \fracm 14 \Psi^{\a}\D^2\Psi_{\a} 
~+ \fracm 12 \Psi^{\a}\Dd^2\Psi_{\a} +c.c.  \big]  ~
\Big\} 
}   \label{equ67}
\ee
This is invariant under the transformation
\be
\delta\Psi_{\a}=\D_{\a}K+i\Dd^2\D_{\a}U     ~~,   \label{equ68}
\ee
with $K=\bar{K}$,~$U=\bar{U}$.  This formulation was considered by Ogievetsky and 
Sokatchev in \cite{Ogievetsky:1975vk} and was noted later in 
\cite{Gates:1979gv} and describes a  massless $Y=1$ supermultiplet.  Referring back to this work, it can be seen that the following
spectrum of fields was presented.
\begin{center}
\begin{tabular} {| c | c | c |}
\hline
Component Field(s) & Bosonic  & Fermionic\\
\hline
$A_{\a \ad}$ & $3$ & {}\\
\hline
$\psi_{\a \b \ad}$~/~$\psi_{\a \ad}$   & {} & $  12  $\\
\hline
$P$ & $1$ & {}\\
\hline
$\lambda_{\ad}$ & {} & $ 4 $\\
\hline
$Y_{\a \ad } $ & $ 8 $ & {}\\
\hline
$t_{\a \b } $ & $6$ & {}\\
\hline
$M$ & $1$ & {}\\
\hline
$N$ & $1$ & {}\\
\hline
$\chi_{\a(s)}$ & {} & $4 $\\
\hline\hline
$~$ & $20$ & $20$\\
\hline
\end{tabular}
\end{center}
\noindent
A brief comparison between this table and the previous reveals a surprise, but a
very satisfying one.

To take the limit of the table at the bottom of page seventeen we begin 
by substituting $s$ $=$ 1.  Upon this substitution, any field with a subscript 
that takes a 0-value means that index does not appear on the field.  For any field
with a subscript that takes a value $<$ 0 means that field does not appear
at all.   When these rules are applied and the value $s$ $=$ 1 is used in the
second and third columns, the two table match perfectly\footnote{We anticipated this
in naming the fields that appear in the expansions on page fifteen.}!

In other words, the matter gravitino multiplet described as ``the (3/2,1) superfield 
of O(2) supergravity'' is the lowest member of the FVdWH-series tower of higher spin multiplets 
of such theories. If this is true that means that there must be a duality between the two theories
in case B) and case C). The answer is yes, these theories are dual to each other and the
duality mechanism is provided by the $s=1$ limit of \eqref{equ35}
\be
\eqalign{ {~~~~~}
S=\int d^8z \Big\{
-&\frac{1}{4}a_1\Psi^{\a}\D^2\Psi_{\a}-\frac{1}{2}a_1\Psi^{\a}\Dd^2\Psi_{\a} +c.c. \cr
+&a_1\Psi^{\a}\Dd^{\ad}\D_{\a}{\Bar{\Psi}}_{\ad}+\[\frac{1}{2}\]a_1\Psi^{\a}\D_{\a}\Dd^{\ad}{\Bar {\Psi}}_{\ad}\cr
-&\[\frac{1}{2}\]a_1\left(\D_{\a}\Psi^{\a}+\Dd_{\ad}\bar{\Psi}^{\ad}\right)B\cr
+&a_1\left(\D_{\a}\Dd^2\Psi^{\a}+\Dd_{\ad}\D^2\bar{\Psi}^{\ad}\right)V\cr
-&\frac{1}{4}a_1BB+\frac{1}{2}a_1V\D^{\g}\Dd^2\D_{\g}V\cr
-&a_1B\left(\D^2V+\Dd^2V\right)\Big\}   ~~~,
}
\ee
which is invariant under the transformations
\be
\eqalign{
&\delta\Psi_{\a}=\D_{\a}\Dd^{\ad}U_{\ad}-\Dd^2\Lambda_{\a}     ~~~,\cr 
&\delta V=\Dd^{\ad}U_{\ad}+\D^{\a}\bar{U}_{\a}   ~~,\cr
&\delta B=-\D^{\a}\Dd^2\Lambda_{\a}-\Dd^{\ad}\D^2\bar{\Lambda}_{\ad}  ~~.
}
\ee
From this point forward there are two choices:
\begin{itemize}
\item Choice 1) We can integrate out the auxiliary superfield B as we did in
the general case and this will give us \eqref{equ61}
\item Choice 2) We can work in a gauge where $B=V=0$ and this will give
(up to a redefinition of the gauge parameter) \eqref{equ67}
\end{itemize}\newpage
$$
\vCent
{\setlength{\unitlength}{1mm}
 \begin{picture}(-20,-140)
  \put(-55,-55){\includegraphics[width=4in]{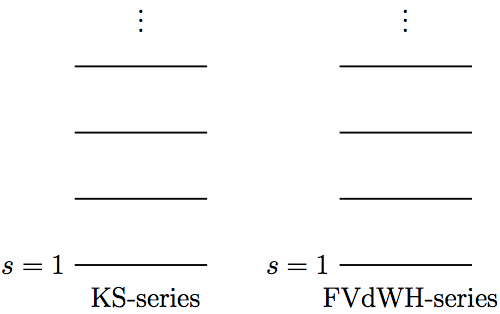}}
\end{picture}}
$$
\vskip2.1in
In our accompanying work of the half-odd integer case \cite{Gates:2011qa},
 we found that the non-minimal
off-shell supergravity theory first discovered by Breitenlohner \cite{nonmin1}, is the
lowest level of an infinite towers of such theories.  In this work we have found
the same thing for ``the (3/2,1) superfield of O(2) supergravity.''  This particular
off-shell matter gravitino multiplet together with the non-minimal off-shell supergravity 
multiplet provides a description of 4D, $\cal N$ $=$ 2 supergravity \cite{deWvanH,
N2SUSYn1SF}.  It is therefore reasonable 
to expect\footnote{A similar realization of 4D, $\cal N$ $=$ 2 supersymmetry has been found
previously \cite{N2}.} that this 4D, $\cal N$ $=$ 2 supersymmetry can
persist when the B-series (${\bm{\cal W}}_{\a(2s +1)}$, $ {\bm{{\cal G}}}_{\a(s)\ad(
s)}$ and ${\bm{{\cal T}}}_{\a(s)\ad(s-1)}$) and
FVdWH-series (${\bm{\cal W}}_{\a(2s)}$, $ {\bm{{\cal P}}}_{\a(s-1)\ad(
s-1)}$ and ${\bm{{\cal T}}}_{\a(s)\ad(s-1)}$) towers are taken together.  A similar behavior
was observed for alternate towers \cite{N2}.
\vskip.2in

${~~~}$ \newline
${~~~~~}$``{\it {The opposite of a correct statement is a false statement. But the  }}${~~~}$ \newline
${~~~~~~~}${\it {opposite 
of a profound truth may well be another profound truth.
}}''
\newline $~~~~~~~$ -- Niels Bohr
\newline ${~~~}$

\noindent
{\Large\bf Acknowledgments}

This research was supported in part by the endowment of the John S.~Toll Professorship, the University of Maryland 
Center for String \& Particle Theory, National Science Foundation Grant PHY-0354401.  This work is also supported 
by U.S. Department of Energy (D.O.E.) under cooperative agreement DEFG0205ER41360.  SJG offers additional 
gratitude to the M.\ L.\ K. Visiting Professorship program and to the M.\ I.\ T.\ Center for Theoretical Physics for support and
hospitality extended during the undertaking of this work.

\end{document}